%% file: main.tex
\documentclass{article}

\usepackage{arxiv}

\usepackage[utf8]{inputenc} %
\usepackage[T1]{fontenc}    %
\usepackage{hyperref}       %
\usepackage{url}            %
\usepackage{booktabs}       %
\usepackage{amsfonts}       %
\usepackage{nicefrac}       %
\usepackage{microtype}      %
\usepackage{lipsum}		%

\usepackage{ulem} 
\usepackage{graphicx}
\usepackage{natbib}
\usepackage{doi}
\usepackage{amsmath}
\usepackage{xcolor}
\usepackage{threeparttable}
\usepackage{subcaption}
\usepackage{pgfplots}
\pgfplotsset{compat=1.17}
\usepackage[capitalize,noabbrev]{cleveref}
\DeclareMathOperator*{\argmin}{arg\,min}
\DeclareMathOperator*{\argmax}{arg\,max}
\usepackage{amssymb}

\title{
    \raisebox{-0.85cm}{\includegraphics[width=6cm]{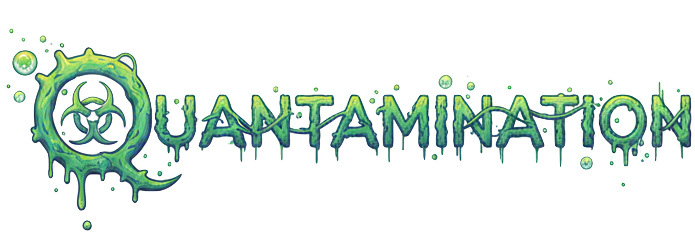}}: Dynamic Quantization Can\\Leak Your Data Across the Batch
}

\date{} 					%
    
\author{
    Hanna Foerster\textsuperscript{1,2} \And
    Ilia Shumailov\textsuperscript{2} \And
    Cheng Zhang\textsuperscript{2} \And
    Yiren Zhao\textsuperscript{2} \AND
    Jamie Hayes\textsuperscript{3} \And
    Robert Mullins\textsuperscript{1}\AND
    \textnormal{\textsuperscript{1}University of Cambridge \quad
    \textsuperscript{2}AI Sequrity Company \quad
    \textsuperscript{3}Google DeepMind} %
}

\renewcommand{\shorttitle}{Dynamic Quantization Leaks Your Data Across the Batch}

\begin{document}
\maketitle

\begin{abstract}
Dynamic quantization emerged as a practical approach to increase the utilization and efficiency of the machine learning serving flow. 
Unlike static quantization, which applies quantization offline, dynamic quantization operates on tensors at run-time, adapting its parameters to the actual input data. Today’s mainstream machine learning frameworks -- including ML compilers and inference engines -- frequently recommend dynamic quantization as an initial step for optimizing model serving. 
This is because dynamic quantization can significantly reduce memory usage and computational load, leading to faster token generation and improved model serving efficiency without substantial loss in model accuracy.
In this paper, we reveal a critical vulnerability in dynamic quantization: an adversary can exploit such quantization strategy to steal sensitive user data placed in the same batch as the adversary’s input. 
Our analysis demonstrates that dynamic quantization, when improperly implemented or configured, can create side channels that expose information about other inputs within the same batch. We call this phenomenon \textit{\textbf{Quantamination}}, describing contamination from quantization. 
Specifically, we show that at least $4$ of the most popular ML frameworks in use today either default to or can use configurations that leak data across the batch boundary. 
This data leakage, in theory, allows attackers to partially or even fully recover other users’ batched input data, representing a serious privacy risk for existing ML serving frameworks.
\end{abstract}

\input{content/01_introduction}
\input{content/02_related_works}
\input{content/03_methodology}

\input{content/04_evaluation}
\input{content/05_discussion}

\section*{Acknowledgements}
We want to thank Nicholas Carlini for providing feedback throughout the project. 

\bibliographystyle{unsrtnat}
\bibliography{references}

\end{document}

%% file: content/01_introduction.tex
\section{Introduction}

Modern LLM serving frameworks such as vLLM \citep{kwon2023vllm}, SGLang \citep{zheng2024sglang}, and DeepSpeed \citep{rasley2020deepspeed} routinely batch concurrent requests from multiple users into shared forward passes to maximize GPU utilization and throughput. To further reduce inference cost and memory footprint, these systems increasingly deploy post-training quantization, converting high-precision activations to lower-precision formats like INT8 or FP8 at runtime. These activation quantization parameters can either be determined statically ahead of time through a calibration set, or computed dynamically during the forward pass for greater accuracy. While many modern serving frameworks compute such dynamic activation quantization parameters for each token separately, several also support computing these parameters across the entire batch.

We find that batching and quantization can create unexpected privacy leakages when configured to compute quantization parameters dynamically across the batch. In this setting, all samples in the batch jointly determine the shared quantization parameters through a global \texttt{min}/\texttt{max} computation over the combined activation tensor, creating a cross-user side channel: the victim's input influences the quantization applied to the adversary's activations, producing measurable differences in the adversary's output. By crafting an input, co-locating it in the same batch as the victim, and observing the resulting perturbation, an adversary can systematically infer properties of the victim's data.

We demonstrate practical attacks exploiting this side channel in both LLM and classification settings, achieving 99.6--100\% token recovery accuracy in LLMs and exact image identification in classification tasks, requiring only batch co-location and observation of top-1 log probabilities. We survey the vulnerability surface across major inference frameworks, identifying at-risk configurations in vLLM's default FP8 online quantization, SGLang's per-tensor FP8 mode, and the dynamic quantization APIs of ONNX Runtime and PyTorch. Notably, per-token dynamic quantization, already the standard in many LLM serving frameworks for both accuracy and computational efficiency reasons, eliminates this side channel entirely, highlighting that per-tensor dynamic activation quantization cannot only be suboptimal for performance but also introduces a concrete privacy vulnerability in multi-tenant serving environments. Finally, we discuss real-world deployment challenges that complicate exploitation in production, including non-determinism from hardware and software heterogeneity and possible provider-side adaptations that conflict with the supposedly deterministic temperature 0 setting.

\begin{figure}[t]
  \centering
  \includegraphics[width=\columnwidth]{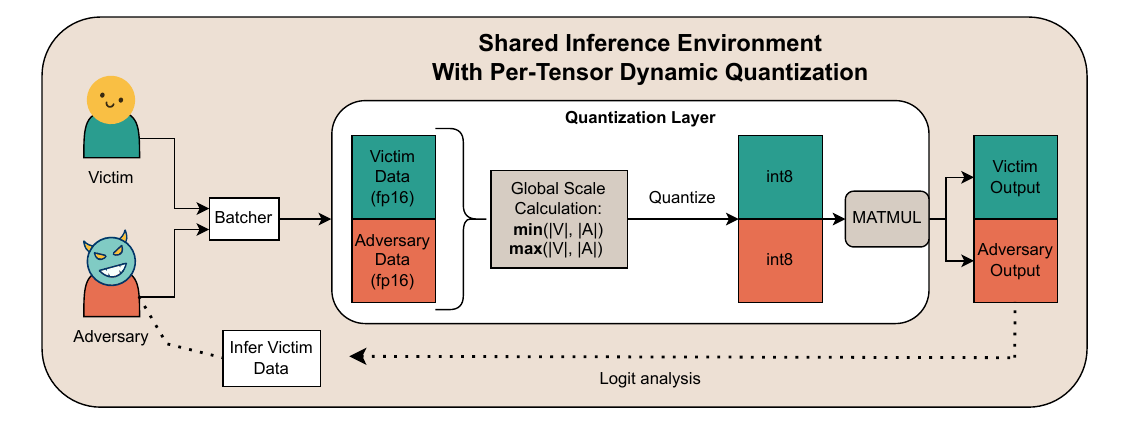}
  \caption{Overview of the per-tensor dynamic activation quantization side channel. A victim and an adversary's inputs are batched together in a shared forward pass. During per-tensor dynamic quantization, a global scale factor is computed from the combined activations of both inputs, causing the victim's data to influence the quantization applied to the adversary's activations. By observing perturbations in its own output logits, the adversary can infer properties of the victim's input.}
  \label{fig:quantization_overview}
\end{figure}

%% file: content/02_related_works.tex
\section{Related Works}

\paragraph{Batch-based privacy attacks.} 

Several recent works demonstrate that shared computation in batched inference creates cross-user information leakage. \citet{yona2024stealing} show that in Mixture-of-Experts (MoE) models, an attacker co-located in the same batch can steal victim prompts by exploiting the tie-handling behavior of Expert-Choice-Routing, and \citet{hayes2024buffer} demonstrate that an adversary can manipulate expert routing by overflowing expert buffers, causing both output degradation and targeted manipulation of co-batched victims' outputs. \citet{kuchler2025architectural} take a complementary approach, demonstrating that architectural backdoors can be embedded into models to enable within-batch data stealing and inference manipulation. \citet{wu2025know} exploit prefix-level KV cache sharing across requests in multi-tenant LLM serving (SGLang, vLLM) to reconstruct prior users' prompts via cache-timing side channels. In federated learning, \citet{yin2021see} demonstrate that batch-averaged gradients can be inverted to recover individual training images, with batch normalization statistics providing additional reconstruction signal. Our attack differs from all of the above in that it requires no architectural modification, no MoE routing, no cache sharing, and is not in a federated learning training setting. Instead it exploits a side channel inherent to inference frameworks that perform per-tensor dynamic quantization, a default or optional configuration in several widely deployed systems.

\paragraph{Numerical error as information channel.} A growing body of work recognizes that numerical artifacts in neural network computation carry exploitable information. \citet{schloegl2021, schloegl2021innformant, schloegl2023} demonstrate that framework- and hardware-specific floating-point deviations serve as forensic fingerprints, enabling identification of the hardware used to produce a given model output. \citet{zhang2025hardware} systematize this observation, showing that numerical differences across platforms are sufficient to reliably infer the hardware and software environment in which a model was executed. \citet{clifford2024locking} exploit similar device-specific numerical behavior to bind models to specific hardware, while \citet{jia2021exploiting} and \citet{zombori2021fooling} show that floating-point error can be weaponized to fool neural network verifiers. In the quantization domain specifically, \citet{matx2026leaky} identify temporal information leakage through block quantization scales during training, and \citet{egashira2024exploiting} show that an adversary can craft weights that activate malicious behavior only after quantization (an integrity attack on the model supply chain). Our work reveals a distinct inference-time privacy risk that requires no adversarial crafting: the data-dependent quantization parameters computed during standard batched inference create exploitable cross-sample information flow, exposing numerical artifacts as a privacy side channel.

\section{Background}

\subsection{Post-Training Quantization}

Post-training quantization reduces the numerical precision of neural network weights and activations from high-precision floating-point (FP32/FP16) to short number formats (INT8, FP8, INT4) to accelerate inference and reduce memory footprint.
The quantization process relies on calibrated parameters to clamp, shift, and scale weights and activations before rounding to low-precision grids to recover model performance.

\textbf{Weight quantization} is usually performed offline before deployment.
Given a trained network,
weights are considered as constants; thus,
complex calibration and quantization algorithms can be applied to yield low-precision weights before deployment.
Modern weight-only approaches like GPTQ~\citep{frantar2022gptq} and AWQ~\citep{lin2024awq} achieve 3--4$\times$ model compression by quantizing weights to INT4/INT8, with per-token/channel calibrated scales to minimize accuracy loss.
Unlike quantization of weights, activation quantization is more challenging 
as the activations change with each input at inference time.
\textbf{Activation quantization} parameters can be determined in two ways: (1)~computed offline from calibration data and fixed for all inputs (\emph{static} activation quantization), or (2)~computed at runtime from the observed values of each forward pass (\emph{dynamic} activation quantization)~\citep{krishnamoorthi2018quantizing, wu2020integer}.
Dynamic quantization introduces more runtime cost compared to static quantization,
but is usually more effective for extremely low precisions like INT4.
Several inference frameworks provide static activation quantization --- for example, TensorRT-LLM's default FP8 and INT8 SmoothQuant modes calibrate per-tensor or per-token activation scales offline and bake them into the model at conversion time~\citep{tensorrt-llm-precision-2024}.
Because these scales do not depend on runtime inputs, static quantization does not create cross-sample information flow and is outside the scope of our attack.
Dynamic activation quantization adapts to input variation and usually achieves higher accuracy, but introduces data-dependent computation of scaling parameters during inference. It is this runtime dependence that our attack exploits.

Existing dynamic activation quantization methods operate at different granularities.
\emph{Per-tensor} quantization computes a single scale factor from the global \texttt{min}/\texttt{max} across an entire activation tensor.
This approach degrades accuracy when activation magnitudes vary significantly, a common occurrence in large language models processing diverse sequences~\citep{dettmers2022gpt3, xiao2023smoothquant}.
\emph{Per-token} quantization addresses this by computing separate scale factors for each token, preserving accuracy for inputs with heterogeneous activation distributions.
Methods like LLM.int8()~\citep{dettmers2022gpt3} and ZeroQuant~\citep{yao2022zeroquant} demonstrate that per-token dynamic quantization maintains model performance at scale.
On modern GPUs, per-token quantization can also be \emph{faster} than per-tensor: computing per-tensor \texttt{min}/\texttt{max} requires a global reduction across all $B \times D$ elements with expensive cross-block synchronization, whereas per-token computes $B$ independent reductions over $D$ elements each, mapping naturally to GPU parallelism without cross-block communication.
Consequently, per-token dynamic quantization has become the standard in LLM-serving frameworks for INT8 precision.
However, per-tensor dynamic quantization persists in two important settings: (1)~FP8 online quantization modes in LLM-serving frameworks such as vLLM~\citep{vllm-fp8-2024}, where it is the \emph{default}; and (2)~general-purpose frameworks such as ONNX Runtime~\citep{onnx-quant-docs-2024} and PyTorch~\citep{pytorch-quant-docs-2024}, whose INT8 dynamic quantization predates the per-token designs adopted by LLM-specific systems.
\textbf{When per-tensor dynamic quantization is applied during batched inference, all samples in the batch contribute to the shared \texttt{min}/\texttt{max} computation, creating cross-batch information flow that enables the privacy attacks demonstrated in this work.}

\subsection{Framework Deployment Landscape}

Table~\ref{tab:framework-landscape} surveys dynamic activation quantization configurations across major inference frameworks.
Frameworks that use static calibrated activation scales (e.g., TensorRT-LLM's default \texttt{fp8} and \texttt{int8\_sq} modes~\citep{tensorrt-llm-precision-2024}, LiteRT's full-integer quantization~\citep{tflite-quant-spec-2024}) embed fixed scales into the model at conversion time and are excluded, as they do not create runtime cross-batch information flow.
Among LLM-serving frameworks, per-token dynamic activation quantization is the standard for INT8: SGLang~\citep{zheng2024sglang}, DeepSpeed-Inference~\citep{yao2022zeroquant},  vLLM~\citep{kwon2023vllm}, and optional dynamic mode of LiteRT~\citep{xnnpack-blog-2024} all default to per-token granularity for their INT8 W8A8 modes.

The vulnerability surface lies primarily in FP8 online quantization, where per-tensor remains the default.
In vLLM~\citep{vllm-fp8-2024}, launching with \texttt{-{}-quantization fp8} dynamically quantizes activations per-tensor at runtime; users must instead use the llm-compressor pipeline~\citep{llmcompressor2024} with \texttt{FP8\_DYNAMIC} to obtain dynamic per-token FP8 scales.
SGLang~\citep{sglang-quant-docs-2025} exposes the choice explicitly through its torchao integration: \texttt{fp8dq-per\_tensor} and \texttt{fp8dq-per\_row} select between the vulnerable and safe granularities respectively, while its \texttt{w8a8\_fp8/int8} and \texttt{int8dq} modes default to per-token.
Beyond LLM-serving, general-purpose frameworks also exhibit the vulnerability.
ONNX Runtime's \texttt{quantize\_dynamic()}~\citep{onnx-quant-docs-2024} computes a single per-tensor activation scale at runtime. Similarly, PyTorch's quantization observer API defaults to per-tensor activation granularity (via \texttt{torch.ao.quantization.observer.PerTensor}); in recent versions, this path shares the same torchao backend used by SGLang's \texttt{fp8dq-per\_tensor} mode.
With our side-channel attack, we highlight that per-tensor dynamic activation quantization is not only harmful for accuracy but also introduces a concrete security vulnerability exploitable across co-located batch samples. In other words, it represents a worst-case scenario: it compromises user privacy while simultaneously degrading model performance.

\definecolor{DarkGreen}{RGB}{1,50,32}
\begin{table}[t]
\centering
\caption{Dynamic activation quantization configurations across major
inference frameworks. Only methods that quantize activations
\emph{at runtime} are shown; weight-only methods (e.g., INT4
AWQ/GPTQ) leave activations in FP16/BF16, and static calibrated
activation scales (e.g., TensorRT-LLM defaults, LiteRT
full-integer quantization) use fixed scales from conversion time---
neither creates runtime cross-batch information flow.
\textbf{Default}: granularity obtained when enabling the
configuration without additional flags;
\textbf{Opt.}: requires selecting a specific flag or variant.
Most LLM-serving frameworks implement per-token dynamic activation
quantization for INT8; per-tensor dynamic quantization appears in
FP8 online modes and in general-purpose frameworks
(ONNX~Runtime, PyTorch).}
\label{tab:framework-landscape}
\small
\setlength{\tabcolsep}{4pt}
\begin{tabular}{@{}lllcl@{}}
\toprule
\textbf{Framework} & \textbf{Configuration} & \textbf{Precision} & \textbf{Setting} & \textbf{Act.\ Gran.} \\
\midrule
\multicolumn{5}{@{}l}{\textit{Per-tensor --- {\color{red}\textbf{vulnerable to cross-batch leakage}}}} \\
\midrule
vLLM~\citep{vllm-fp8-2024}
  & \texttt{-{}-quantization fp8} (online)
  & W8A8 FP8 & Default & Per-tensor \\
SGLang~\citep{sglang-quant-docs-2025}
  & \texttt{-{}-torchao-config fp8dq-per\_tensor}
  & W8A8 FP8 & Opt. & Per-tensor \\
ONNX RT~\citep{onnx-quant-docs-2024}
  & \texttt{quantize\_dynamic()}
  & W8A8 INT8 & Default & Per-tensor \\
PyTorch~\citep{pytorch-pertensor-observer-2025}
  & {\footnotesize\texttt{torch.ao.quantization...PerTensor}}
  & W8A8 FP8/INT8 & Default & Per-tensor \\
\midrule
\multicolumn{5}{@{}l}{\textit{Per-token / per-row --- {\textbf{\color{DarkGreen}not vulnerable}}}} \\
\midrule
SGLang~\citep{sglang-quant-docs-2025}
  & \texttt{-{}-torchao-config fp8dq-per\_row}
  & W8A8 FP8 & Opt. & Per-token \\
SGLang~\citep{sglang-quant-docs-2025}
  & \texttt{-{}-torchao-config int8dq}
  & W8A8 INT8 & Default & Per-token \\
SGLang~\citep{sglang-quant-docs-2025}
  & \texttt{-{}-quantization w8a8\_fp8/int8}
  & W8A8 FP8/INT8 & Default & Per-token \\
vLLM~\citep{kwon2023vllm}
  & W8A8 FP8/INT8 (llm-compressor)
  & W8A8 FP8/INT8 & Default & Per-token \\
DeepSpeed~\citep{yao2022zeroquant}
  & ZeroQuant W8A8
  & W8A8 INT8 & Default & Per-token \\
\bottomrule
\end{tabular}
\end{table}

%% file: content/03_methodology.tex
\section{Methodology}

\subsection{Threat Model}
\label{sec:threat-model}

We consider an adversary who seeks to infer sensitive information about a victim's input by exploiting side-channels introduced by dynamic quantization in batched inference settings.

\paragraph{Notation.}
Let $s$ denote the victim's secret input, $a$ denote the adversarial input controlled by the attacker, and $C = \{c_1, \ldots, c_k\}$ represent a candidate set of possible victim inputs. We use $\text{logits}(a, x)$ to denote the model's logit output for input $a$ when batched with input $x$. For LLM token recovery, we denote the victim's token sequence as $s = [s_1, s_2, \ldots, s_n]$ and the adversary's token sequence as $a = [a_1, a_2, \ldots, a_m]$, where $s_i, a_j \in V$ and $V$ is the model's vocabulary.

\paragraph{Attack surface.}
In the per-tensor setting, dynamic quantization computes a single set of batch-specific quantization parameters (scale and zero-point) based on the activation statistics of all inputs within a batch. Since these parameters are shared across the entire batch, the adversary's injected input directly influences the quantization of the victim's data. By strategically crafting an adversarial input and observing the resulting change in their own output, the adversary can infer properties of the victim's input.

\paragraph{Attack goal.}
The adversary aims to recover the victim's token sequence $s = [s_1, \ldots, s_n]$ (LLM tasks) or identify the class/find similar candidates to the victim's input $s$ (classification tasks).

\paragraph{Adversary capabilities.}
The adversary possesses the following capabilities:
\begin{enumerate}
    \item \textbf{Batch co-location:} The adversary can inject a data point $a$ into a batch that is processed jointly with the victim's input $s$. For simplicity, we assume a batch size of $2$, where one data point is from the adversary and the other from the victim.
    \item \textbf{Output observation:} The adversary can observe the model's output (logits or predictions) corresponding to their own injected data points.
    \item \textbf{Candidate set access:} 
        \begin{itemize}
            \item \textit{LLM setting:} The adversary knows the vocabulary $V$.
            \item \textit{Classification setting:} The adversary has access to a public dataset from the same distribution as the victim's data point. We assume two settings, where either the dataset contains the victim's data point or does not contain the victim's data point.
        \end{itemize}
    \item \textbf{Model knowledge:} The adversary knows which model is being used and knows that dynamic quantization is employed during inference.
    \item \textbf{Model access:}
    \begin{itemize}
        \item \textit{White-box (open-source):} The adversary can run the model locally.
            \begin{itemize}
                \item LLM setting: Enables testing candidates with arbitrary KV cache states.
                \item Classification setting: Enables an optimized adversarial probe.
            \end{itemize}
        \item \textit{Black-box (closed-source):} The adversary queries via API only. This adds the assumption that the adversary is able to fill many batches with only their own data points when testing with adversary and candidate data points.
            \begin{itemize}
                \item LLM setting: Requires greedy decoding assumption for reproducible adversarial sequences.
                \item Classification setting: Limits to random adversarial probe selection.
            \end{itemize}
    \end{itemize}
\end{enumerate}

\paragraph{Threat model justification.} \textit{Batch co-location} arises naturally in multi-tenant serving systems that batch concurrent requests for throughput optimization. For experiments in the paper we use a batch size of 2 to simplify analysis and clearly isolate quantization effects. In practice, an attacker for a batch size $N$ can submit $N-1$ items that are the same, thus effectively reducing the setting to our 2-item batch setup. Larger batch sizes in production systems may add more noise to these quantization effects but would not eliminate the core vulnerability, since any co-located inputs can influence shared quantization parameters. \textit{Output observation} is a realistic assumption, as APIs like OpenRouter expose parameters like top-k logprobs. We assume access to top-1 logprobs for the LLM case and the full logit vector for the classification case, as we only test on a 10-class problem. \textit{Candidate set access} is inherent for LLMs as tokenizers are publicly available for all major models. For classification, we evaluate two scenarios: (1) the victim's input exists in a public dataset from the same distribution, representing a threat where the adversary identifies which specific datapoint the victim queried, and (2) the victim's input is excluded from the candidate set, where the adversary seeks similar inputs or infers the victim's class label. \textit{Model knowledge} is reasonable as adversaries typically initiate queries and can infer model details from API behavior or documentation. For \textit{Model access}, white-box attacks on open-source models (LLaMA, Mistral, Qwen) are highly practical given the prevalence of locally-runnable models and represent the strongest adversarial capability. Black-box attacks on closed APIs require additional assumptions (e.g., greedy decoding for LLMs, random probe selection for classification) but still constitute a meaningful threat model.

\subsection{Dynamic Quantization Mechanics in Batched Inference}

\paragraph{Quantization parameter computation}
Dynamic INT8 quantization converts floating-point activations to integers at inference time.
Given an input batch $\mathbf{X} \in \mathbb{R}^{b, d}$ where $b$ denotes batch size and $d$ denotes hidden dimension,
the per-batch scale $\alpha$ and the zero point $\beta$ are calculated as
\begin{equation}
   \alpha = \frac{ \max{(\mathbf{X}) - \min{(\mathbf{X})}}}{255}\;,
   \beta = -\mathrm{round}(\frac{\min{(\mathbf{X})}}{\alpha}).
\end{equation}
where $\max{(\cdot)}$ and $\min{(\cdot)}$ calculate the maximum and minimum values across the entire matrix, and $\mathrm{round}{(\cdot)}$ denotes elementwise rounding to the nearest (RTN).

The input batch is then quantized on the fly:
\begin{equation}
    \mathbf{X}_q = \mathrm{clip}\left( \mathrm{round}( \frac{\mathbf{X}}{\alpha} + \beta),\, 0,\, 255 \right)
\end{equation}
where $\mathrm{clip}(\cdot)$ clamps every element to the INT8 range (0-255).

\textbf{Key observation:} An input sample with a larger activation magnitude dominates quantization parameters ($\alpha$ and $\beta$) of a batch. Different co-batched inputs create measurably different quantization effects, observable in final logits through $\Delta(a,s,c) = \|\text{logits}(a,s) - \text{logits}(a,c)\|$. In multi-layer networks, per-layer quantization effects compound in complex ways (discussed further in Section~\ref{sec:classification-attack}).

\subsection{Attack Methodology: LLM Token Recovery}
\label{sec:llm-attack}

The LLM setting is particularly vulnerable because the vocabulary $V$ is known and fixed. We recover the secret token sequence $s = [s_1, \ldots, s_n]$ iteratively, exploiting batching behavior during the decode phase of LLM inference.

\paragraph{Batching in LLM inference.}
LLM inference consists of two phases with distinct batching characteristics. The \textit{prefill phase} processes the entire prompt in parallel, where multiple tokens from the same sequence (or multiple sequences) can be batched together, creating complex quantization interactions. The \textit{decode phase} generates tokens sequentially, one per forward pass, where each forward pass in batched serving typically processes one new token per active sequence. We target the decode phase where each step processes exactly two tokens, the adversarial token $a_i$ and the victim's token $s_i$, with shared quantization parameters, creating a direct observable side-channel.

\paragraph{Batching persistence assumption.}

Modern LLM serving systems (e.g., vLLM, TensorRT-LLM) use continuous batching where sequences are grouped together and remain co-batched throughout their generation. Once the adversary's sequence and victim's sequence are batched together in the decode phase, they continue processing in the same batch until one completes. This persistence is realistic because re-batching mid-generation would require expensive KV cache reorganization, which serving systems avoid for efficiency.

\textbf{Token-by-token recovery algorithm.}

\textit{Initial token recovery ($i = 1$):}

For the first token, we have no context to constrain the search space, requiring broader exploration:
\begin{enumerate}
    \item \textit{Query candidate tokens:} For each $t^{(j)} \in V$ measure:
    \begin{equation}
        \Delta_1(t) = \|\text{logits}(a_1, s_1) - \text{logits}(a_1, [t^{(j)}])\|_2.
    \end{equation}
    
    \item \textit{Match with early termination:} Return the first token satisfying:
    \begin{equation}
        \hat{s}_1 = \argmin_{t \in V} \Delta_1(t) \quad \text{subject to} \quad \Delta_1(\hat{s}_1) < \epsilon,
    \end{equation}
    where $\epsilon$ is a match threshold.
    
    \item \textit{Optional: Prior-based ordering:} To reduce queries, order candidates by sentence-initial probability $p_{\text{init}}(t)$ before testing. Tokens like \texttt{<BOS>}, \texttt{"The"}, \texttt{"I"} are tested first, exploiting statistical biases in natural language.
\end{enumerate}

\textit{Subsequent token recovery ($i > 1$):}

For later tokens, context $s_{<i} = [s_1, \ldots, s_{i-1}]$ provides strong constraints, enabling efficient search:
\begin{enumerate}
    \item \textit{Rank by language model prior:} Compute $p(t | s_{<i})$ for all $t \in V$ and sort in descending order of probability.
    
    \item \textit{Sequential testing:} Iterate through ranked candidates $t^{(1)}, t^{(2)}, \ldots$ where $p(t^{(j)} | s_{<i}) \geq p(t^{(j+1)} | s_{<i})$:
    \begin{equation}
        \Delta_i(t^{(j)}) = \|\text{logits}(a_i, s_i) - \text{logits}(a_i, t^{(j)})\|_2.
    \end{equation}
    
    \item \textit{Early termination:} Stop at the first candidate satisfying:
    \begin{equation}
        \hat{s}_i = t^{(j)} \quad \text{if} \quad \Delta_i(t^{(j)}) < \epsilon.
    \end{equation}
\end{enumerate}

Depending on threat model, this recovery algorithm needs to be interpreted with slightly different assumptions.

\textbf{White-box setting (open-source model):} The adversary runs the model locally for candidate testing. For testing candidate tokens $t^{(j)} \in V$ at position $i$, the adversary prefills the KV cache with previously found secret sequence context $\hat{s}_{<i}$ and uses the actual adversarial token sequence $[a_1, \ldots, a_i]$ observed during victim interaction to measure $\Delta_i(t^{(j)})$. No assumptions about the adversary's decoding strategy are needed.

\textbf{Black-box setting (closed-source model):} Candidate search must be performed through API queries. To replicate the adversary's token sequence $[a_1, \ldots, a_i]$ for candidate testing, we assume the adversary used greedy decoding (selecting $\argmax$ token at each step). This allows deterministically reconstructing the adversarial sequence state for testing each candidate. Without this assumption, the adversary would need to re-sample their sequence multiple times, hoping to randomly recreate the exact same token sequence $[a_1, \ldots, a_i]$ observed during victim interaction. The probability of matching this sequence decreases exponentially with sequence length, particularly when sampling strategies (temperature, top-k, nucleus sampling) frequently select lower-probability tokens.

\subsection{Attack Methodology: Classification Setting}
\label{sec:classification-attack}

The classification setting is more challenging because exact matches for $s$ may not exist in the adversary's candidate set $C$. Unlike the discrete, finite vocabulary in LLMs, the space of possible images is effectively infinite in practice. In our evaluation, we test two scenarios: (1) a setting where the entire test set (including the secret) is used as the candidate set, representing cases where the adversary aims to identify which specific image from a known dataset the victim queried, and (2) a setting where all test samples except the secret are used as candidates, representing a more realistic scenario where the adversary cannot enumerate all possible images and the victim's input may not exist in any available dataset. As in the LLM token recovery, we assume that the secret input produces a distinct dynamic quantization effect on the adversary's output logit calculation. In the first scenario, we test whether this effect can pinpoint the exact secret if it exists in the candidate set. In the second scenario, we test whether inputs from the same class produce more similar quantization effects than inputs from different classes, allowing class-level inference even without an exact match.

\paragraph{Attack strategy.}

For each candidate $c_j \in C$ and adversarial probe $a$, we compute:
\begin{equation}
    \text{score}(c_i) = \|\text{logits}(a, s) - \text{logits}(a, c_i)\|_2.
\end{equation}

The best candidate is:
\begin{equation}
    \hat{c} = \argmin_{c_i \in C} \text{score}(c_i).
\end{equation}

In multi-layer networks, each layer applies dynamic quantization independently based on that layer's activation statistics. At each layer, the quantization parameters (scale, zero-point) are determined by the minimum and maximum activations in the batch. Either one input sets both extremes (dominating that layer), or each input sets one extreme (split control). Crucially, which input dominates varies layer by layer: input $a$ might set the max at layer 1 but the min at layer 2. These per-layer effects can partially cancel when propagated through the network, making aggregate final logits ambiguous. However, as network depth increases, both the probability that one input dominates across all layers and the probability that per-layer effects cancel out perfectly decrease exponentially, making coincidental aggregate matches increasingly unlikely. Hence, in a significantly deep network, for a candidate $c$ to produce matching logits for an arbitrary adversarial probe $a$, $c$ must exhibit nearly identical per-layer quantization behavior to $s$.

\textbf{White-box setting:} With local model access, the adversary can optimize probe selection by computing per-layer activation ranges for potential probes and selecting one with particularly small activation ranges across multiple layers, maximizing its influence on quantization parameters.

\textbf{Black-box setting:} Limited to API queries, the adversary uses random selection for both adversarial probes and candidates from the available dataset, relying solely on aggregate logit scoring.

%% file: content/04_evaluation.tex
\section{Evaluation}

\subsection{LLM Token Recovery}

\begin{table}[h]
\centering
\caption{Language models and tokenizers used for token recovery evaluation.}
\label{tab:llm-models}
\begin{tabular}{lllll}
\toprule
\textbf{Model} & \textbf{Parameters} & \textbf{Vocab Size} & \textbf{Tokenizer}  \\
\midrule
TinyStories-1M~\citep{eldan2023tinystories} & 1M & 50,257\textsuperscript{*} & GPT-2 Tokenizer~\citep{radford2019language}  \\
Pythia-70M~\citep{biderman2023pythia} & 70M & 50,257 & GPT-NeoX-20B Tokenizer~\citep{black2022gpt}  \\
SmolLM2-135M~\citep{allal2025smollm2smolgoesbig} & 135M & 49,152 & SmolLM tokenizer~\citep{allal2024smollm}  \\
\bottomrule
\end{tabular}
\begin{tablenotes}
\small
\item[*] *TinyStories models use only the top 10,000 most frequent tokens during training, though the full GPT 2/GPT Neo vocabulary contains 50,257 tokens.
\end{tablenotes}
\end{table}

We evaluate the token recovery attack on three small language models with varying architectures and tokenizers: TinyStories-1M~\citep{eldan2023tinystories}, Pythia-70M~\citep{biderman2023pythia}, and SmolLM2-135M~\citep{allal2025smollm2smolgoesbig}. Table~\ref{tab:llm-models} summarizes the model specifications. All experiments use batch size two during the decode phase, with one token from the adversary's sequence and one token from the victim's sequence per forward pass.

The diversity in tokenizer vocabulary sizes and training corpora allows us to assess whether the attack generalizes across different tokenization schemes. TinyStories uses the GPT-2 tokenizer~\citep{radford2019language} (50,257 tokens) 
similar to GPT-Neo~\citep{gpt-neo}, Pythia uses the GPT-NeoX-20B tokenizer~\citep{black2022gpt} trained specifically on The Pile~\citep{gao2020pile}, and SmolLM2 uses a custom tokenizer~\citep{allal2024smollm} trained on the Smollm Corpus.

We sample secret sequences of up to 20 tokens from various datasets and measure the number of candidate tokens evaluated before finding the correct match. All evaluations were run on Intel Xeon Ice Lake CPUs and to ensure practical efficiency, we impose a 4-hour timeout per sequence. Runs exceeding this limit were excluded from the analysis (affecting a small percentage of sequences, especially for out-of-distribution cases). To create a more realistic and challenging attack scenario, we convert the full logit vectors to log probabilities and extract only the top-1 prediction (argmax) rather than using the complete logit distribution. This single scalar value per forward pass represents what an attacker could at most observe from typical model serving APIs. We set the matching threshold to $\epsilon = 10^{-6}$ for the L2 distance between these top-1 log probability values, which we found empirically provides a robust threshold.

\begin{table}[t]
\centering
\caption{Token recovery attack results across models and datasets. All models achieve near-perfect recovery with varying efficiency based on model-data distribution alignment. All settings were run 100 times with a 4-hour timeout per sequence. The Runs column reports how many of 100 trials completed within the 4-hour timeout; incomplete runs are excluded from accuracy and query counts. This highlights the inefficiency of recovering out-of-distribution sequences compared to in-distribution sequences. }
\label{tab:llm-results}
\begin{tabular}{llrrr}
\toprule
\textbf{Model} & \textbf{Dataset} & \textbf{Runs} & \textbf{Accuracy} & \textbf{Mean Queries} \\
\midrule
\multicolumn{5}{l}{\textit{Specialized model}} \\
\ \ TinyStories-1M & TinyStories & 100 & 100.0\% & 424 \\
\ \ TinyStories-1M & WikiText & 12 & 99.6\% & 8,095 \\
\ \ TinyStories-1M & AG News & 8 & 100.0\% & 7,489 \\
\midrule
\multicolumn{5}{l}{\textit{General-purpose models}} \\
\ \ Pythia-70M & WikiText & 96 & 99.6\% & 1,343 \\
\ \ SmolLM2-135M & WikiText & 88 & 100.0\% & 985 \\
\midrule
\multicolumn{5}{l}{\textit{Theoretical baseline (random search)}} \\
\ \ All models & Any & --- & --- & $|V|$/2 $\sim$ 25,000 \\
\bottomrule
\end{tabular}
\end{table}

\paragraph{Attack success.}
Table~\ref{tab:llm-results} summarizes our results. The attack achieves near-perfect token recovery across all settings, with 99.6-100\% accuracy. With our chosen threshold $\epsilon = 10^{-6}$, we observe occasional false positives (0.4\% of tokens for Pythia-70M, 0\% for others). These occur when two different tokens produce nearly identical quantization effects across layers. Stricter thresholds might be able to eliminate these errors.

\paragraph{Attack efficiency and distribution alignment.}
The attack significantly outperforms random search, which would require $|V|/2 \approx 25{,}000$ queries per token on average. In realistic attack scenarios, victim secrets would be generated by the model itself during decoding, ensuring in-distribution alignment. However, to understand how distribution alignment affects attack efficiency, and to evaluate the attack on more meaningful sequences than randomly sampled tokens, we deliberately tested recovery across different dataset sources. This reveals that efficiency varies dramatically based on the alignment between model training distribution and secret text distribution. 

TinyStories-1M, a model specialized for simple children's stories, illustrates this effect. For in-distribution secrets from its native domain, the attack achieves exceptional efficiency (424 queries/token on average, 59$\times$ faster than random). However, for out-of-distribution datasets like WikiText and AG News, efficiency degrades severely to 7,500-8,000 queries per token, with many sequences timing out after 4 hours. While such out-of-distribution scenarios are unrealistic in practice, this comparison demonstrates how critical language priors are to the computational feasibility of this attack. For general-purpose models trained on diverse web text (Pythia-70M and SmolLM2-135M), the attack maintains practical efficiency (985-1,343 queries/token) on WikiText, achieving 18-25$\times$ speedup over random search and demonstrating the general computational feasibility of this attack. We expect that sequences generated directly by these models (the realistic threat scenario) would be even more efficient to recover, as they would exhibit stronger alignment with the model's learned distribution.

\input{content/figures/token_position_scaling}

\paragraph{Context-based acceleration.}
Figure~\ref{fig:position-efficiency} demonstrates how autoregressive context dramatically accelerates recovery. The first token (no context) requires 5,000-7,000 queries across all models, but efficiency improves exponentially at subsequent positions as each recovered token constrains the next. TinyStories-1M exhibits the most dramatic improvement on in-distribution data: after the costly first token ($\sim$4,900 queries), positions 1-19 average only 20-65 queries, reflecting its strong priors over familiar story structures. In contrast, general-purpose models (Pythia-70M, SmolLM2-135M) maintain more consistent mid-range efficiency (200-2,500 queries per position), reflecting broader but less specialized language knowledge. Despite these differences in pattern, all models vastly outperform random search ($\sim$25,000 queries) at every position, with the advantage growing as context accumulates.

\subsection{Classification Model Class Recovery}

\subsubsection{Experimental Setup}
We evaluate the classification attack on MNIST~\citep{lecun1998mnist}, a 10-class handwritten digit classification benchmark with 10,000 test images (1,000 per class). We test three CNN architectures with increasing depth: a simple 3-layer CNN (3 convolutional layers + 2 fully connected layers), ResNet18~\citep{he2016deep} (18 layers), and ResNet50~\citep{he2016deep} (50 layers). All models are trained to convergence on MNIST and achieve $>$98\% test accuracy. We apply 8-bit dynamic quantization during inference using the same batching setup as the LLM experiments (batch size 2: one adversarial probe \textit{a}, one victim input \textit{s}).

\subsubsection{Results: Secret in Candidate Set}

\begin{table}[h]
\centering
\caption{Exact recovery success rate when the secret exists in the candidate set. Each model is evaluated 10 times per probe strategy (20 trials total per model).}
\label{tab:classification-exact}
\begin{tabular}{lcc}
\toprule
\textbf{Model} & \textbf{Random Probe} & \textbf{Layer-Diverse Probe }\\
\midrule
Simple CNN (3 layers) & 8/10 (80\%) & 10/10 (100\%) \\
ResNet18 (18 layers) & 10/10 (100\%) & 10/10 (100\%) \\
ResNet50 (50 layers) & 10/10 (100\%) & 10/10 (100\%) \\
\bottomrule
\end{tabular}
\end{table}

Table~\ref{tab:classification-exact} shows near-perfect exact recovery when the secret exists in the candidate set. For deeper networks (ResNet18, ResNet50), even randomly selected probes achieve 100\% success, confirming that per-layer quantization effects are unlikely to coincidentally cancel across many layers. The simple CNN occasionally fails with random probes (80\% success) because strong random probes can dominate quantization parameters, obscuring the victim's signal. Optimized layer-diverse probes with minimal activation ranges restore perfect recovery. This demonstrates that probe selection matters for shallow networks but becomes less critical as network depth increases.

\subsubsection{Results Excluding Secret From Candidate Set}

\begin{figure}[t]
\centering
\includegraphics[width=0.32\textwidth]{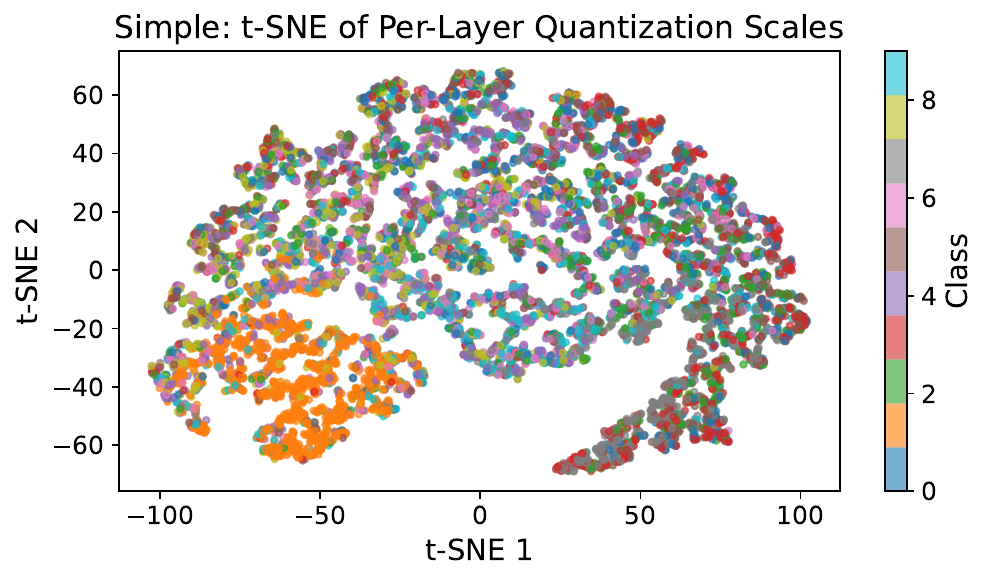}\hfill
\includegraphics[width=0.32\textwidth]{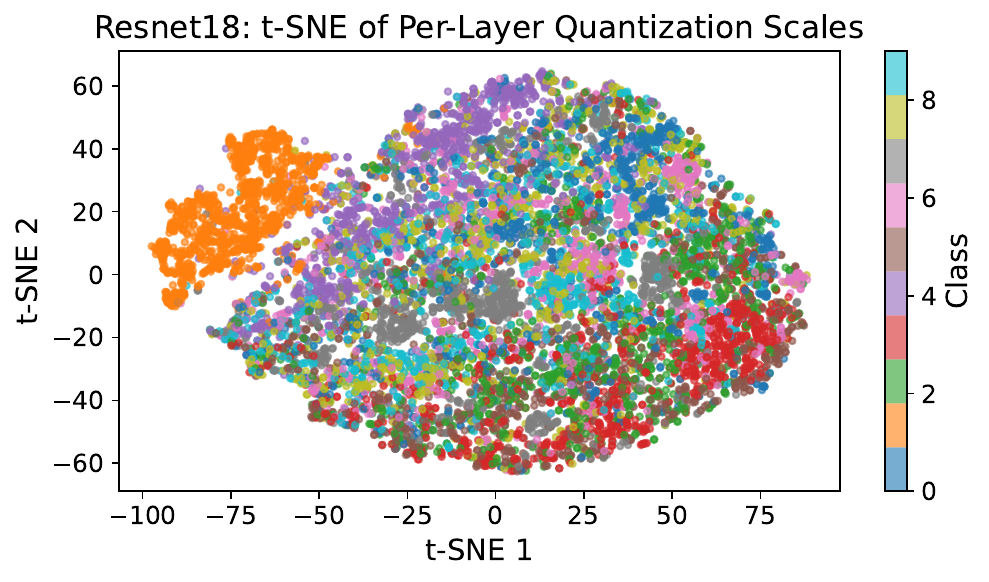}\hfill
\includegraphics[width=0.32\textwidth]{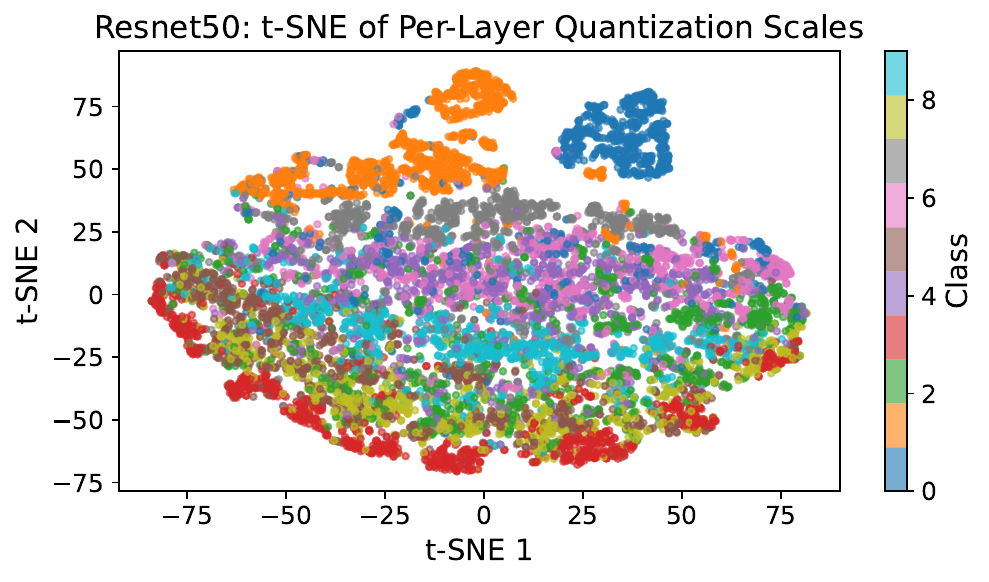}
\vspace{0.2cm}
\includegraphics[width=0.32\textwidth]{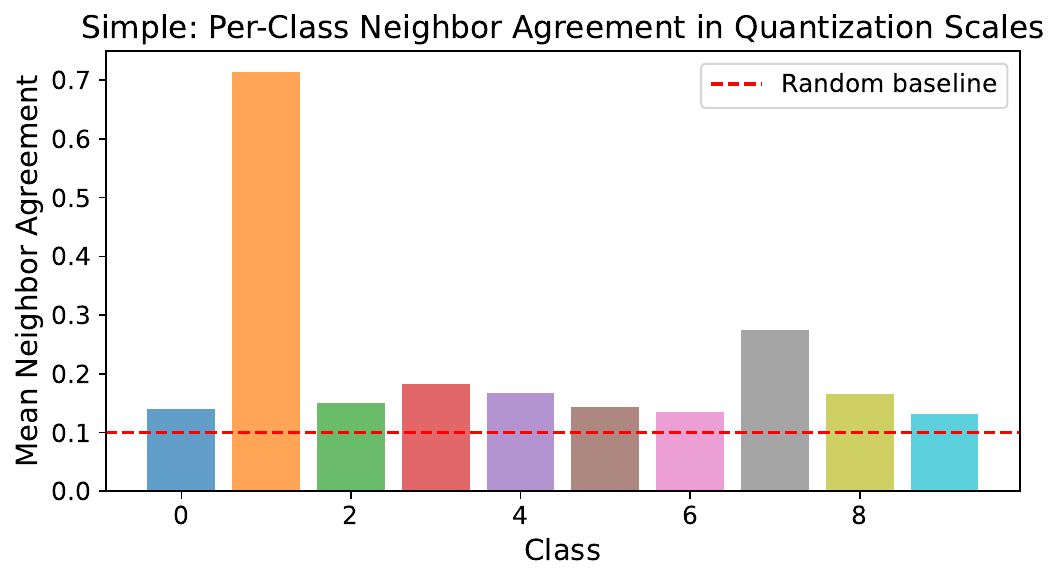}\hfill
\includegraphics[width=0.32\textwidth]{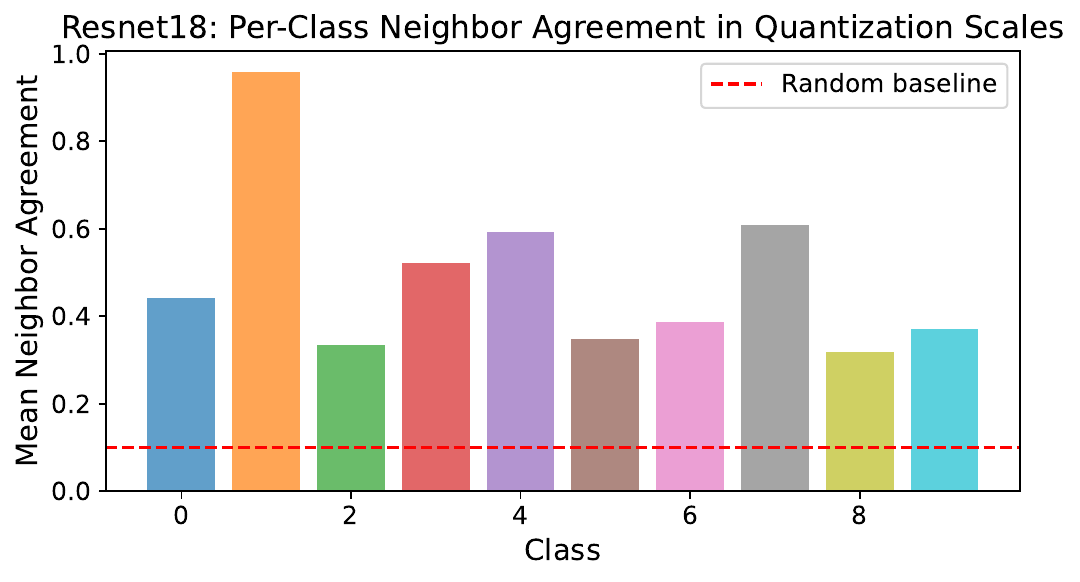}\hfill
\includegraphics[width=0.32\textwidth]{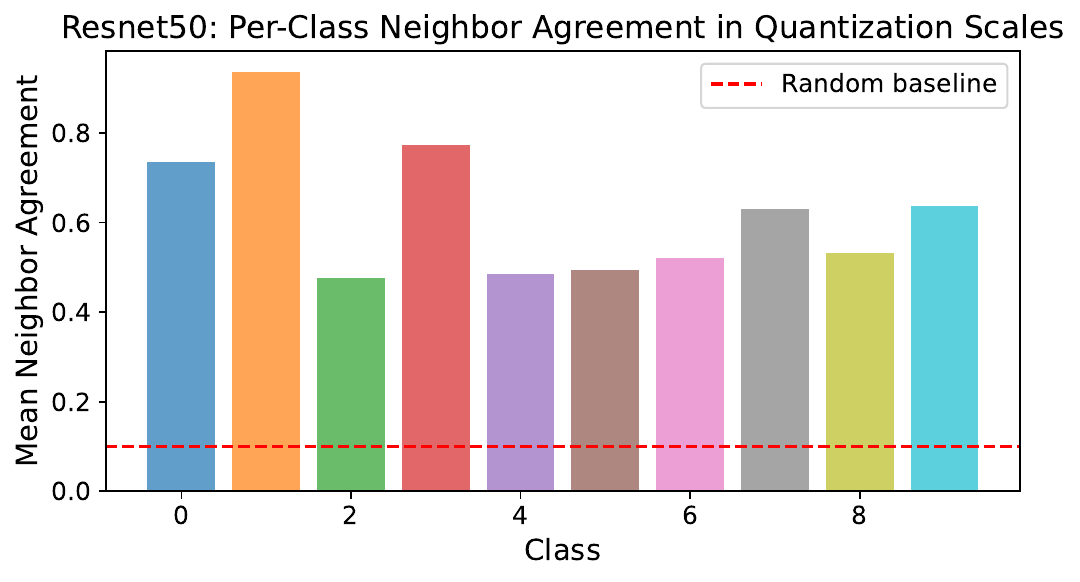}
\caption{\textbf{From left to right:} Analysis of a simple 3-layer CNN, Resnet18, and Resnet50's per-layer dynamic quantization scales aggregated together. \textbf{Upper row:} 2D t-SNE projection of each test sample's per-layer quantization scales, colored by true class. Distinct clusters indicate that different classes induce separable quantization behavior across layers. \textbf{Lower row:} For each class, the average proportion of a sample's 10 nearest neighbors that belong to the same class, measured by similarity of per-layer quantization scales. Bars above the random baseline (red) indicate that samples with similar quantization profiles tend to share a class.}
\label{fig:class_scale_analysis}
\end{figure}

When the secret is excluded from the candidate set (9,999 candidates remaining), exact recovery is impossible. We instead evaluate whether class-level inference is feasible.

\paragraph{Evidence for class-recoverable quantization signatures.}
Figure~\ref{fig:class_scale_analysis} suggests that class recovery may be possible. The upper row shows 2D t-SNE projections of each test sample's per-layer quantization scales, colored by true class. Clustering becomes more pronounced as network depth increases. While the simple CNN shows weak separation, ResNet18 exhibits clearer class structure particularly for class 1, and ResNet50 shows the most distinct clustering separating most distinctly classes 0, 1, and 2. Although overlap remains even for ResNet50, the lower row demonstrates meaningful class correlation: for an average sample, over 50\% of its 10 nearest neighbors (by quantization signature similarity) belong to the same class, well above the 10\% random baseline. This indicates that per-layer quantization scales do carry class information in deep networks.

\paragraph{Attack performance and limitations.}
Despite this promising signal, our attack achieves limited success at class inference. We use optimized layer-diverse probes (to exclude poorly-performing random probes) and evaluate 50 runs per model. We test multiple ranking strategies: top-1 candidate selection, top-3 candidate selection, and clustering-based approaches where we cluster the top 100 candidates into 10 groups by scale profile similarity and select either the closest cluster on average or the cluster containing the top-3 candidates. None of these strategies significantly improve performance. For top-1 candidate accuracy, we achieve 16\% for the simple CNN, 14\% for ResNet18, and 10\% for ResNet50, only marginally above the 10\% random baseline, despite the strong class clustering visible in the scale profiles.

\paragraph{Why class inference fails.}
We hypothesize this failure stems from a fundamental information asymmetry. The clustering analysis in Figure~\ref{fig:class_scale_analysis} directly examines each image's per-layer quantization scale profile. However, during the attack, we cannot observe the secret's scale profile directly. We only observe the difference in logit output it produces when batched with our adversarial probe. This indirect signal limits our ability to locate the correct cluster of candidates with similar scale profiles. Even when classes form distinct clusters in scale space, the logit-based side-channel may not reliably map to those clusters. Future work might explore alternative attack strategies that better exploit the class structure present in quantization signatures.

%% file: content/figures/token_position_scaling.tex
\begin{figure}[t]
\centering
\begin{tikzpicture}
\begin{axis}[
    width=0.9\columnwidth,
    height=6cm,
    xlabel={Token Position},
    ylabel={Mean Candidates Tried (log)},
    ymode=log,
    xmin=0, xmax=19,
    ymin=1, ymax=300000,
    grid=major,
    legend pos=north east,
    legend style={font=\small},
    mark size=1.5pt,
    line width=1pt,
]

\addplot[color=blue, mark=*, mark options={fill=blue}] coordinates {
    (0, 4892.9)
    (1, 23.3)
    (2, 65.5)
    (3, 26.6)
    (4, 43.5)
    (5, 64.8)
    (6, 29.9)
    (7, 19.9)
    (8, 19.0)
    (9, 29.5)
    (10, 53.1)
    (11, 13.6)
    (12, 17.9)
    (13, 42.4)
    (14, 46.4)
    (15, 53.6)
    (16, 47.0)
    (17, 93.6)
    (18, 2.5)
    (19, 1.0)
};
\addlegendentry{TinyStories-1M (in-dist)}

\addplot[color=red, mark=square*, mark options={fill=red}] coordinates {
    (0, 6695.0)
    (1, 2575.5)
    (2, 1650.0)
    (3, 1645.5)
    (4, 1339.6)
    (5, 825.7)
    (6, 1200.2)
    (7, 583.1)
    (8, 387.0)
    (9, 425.3)
    (10, 633.7)
    (11, 462.7)
    (12, 1038.9)
    (13, 540.2)
    (14, 548.9)
    (15, 404.4)
    (16, 824.2)
    (17, 1668.2)
    (18, 819.2)
    (19, 1467.6)
};
\addlegendentry{Pythia-70M}

\addplot[color=green!60!black, mark=triangle*, mark options={fill=green!60!black}] coordinates {
    (0, 5770.6)
    (1, 1620.8)
    (2, 2470.6)
    (3, 1245.3)
    (4, 727.4)
    (5, 223.5)
    (6, 129.9)
    (7, 241.0)
    (8, 878.5)
    (9, 517.4)
    (10, 221.7)
    (11, 484.1)
    (12, 344.0)
    (13, 683.1)
    (14, 100.4)
    (15, 297.7)
    (16, 77.6)
    (17, 362.3)
    (18, 621.1)
    (19, 193.2)
};
\addlegendentry{SmolLM2-135M}

\addplot[color=black, dashed, thick, domain=0:19, samples=2] {25129};
\addlegendentry{Random baseline}

\end{axis}
\end{tikzpicture}
\caption{Query efficiency improves dramatically with position as autoregressive context accumulates. The first token (no context) requires thousands of queries, while subsequent tokens leverage language priors to narrow the search space. TinyStories-1M shows the strongest efficiency gain on in-distribution data, while general-purpose models (Pythia, SmolLM2) maintain consistent mid-range efficiency. All models vastly outperform random search baseline ($\sim$25,000 queries).}
\label{fig:position-efficiency}
\end{figure}

%% file: content/05_discussion.tex
\section{Discussion}
\label{sec:discussion}

\begin{figure}[t]
\centering
\includegraphics[width=0.95\textwidth]{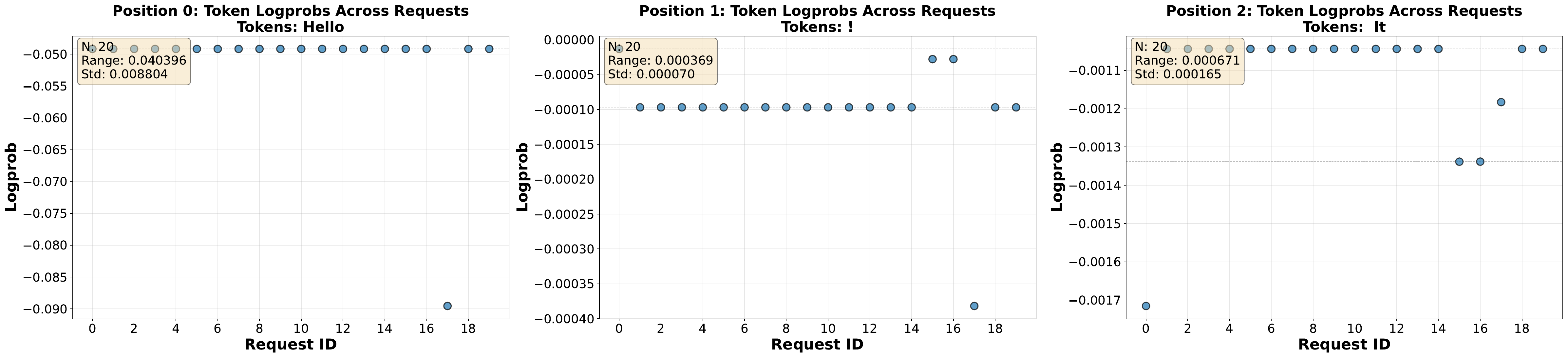}
\caption{Llama Log prob analysis of the first 3 output tokens after querying "Hello my name is" 20 times with temperature 0 in OpenRouter with provider Crusoe. }
\label{fig:llama_logprobs}
\end{figure}

\textbf{Real-world deployment challenges: Non-determinism. }Our attack relies on detecting small logit differences caused by cross-batch quantization effects, which raises the question of whether such signals survive the noise present in production deployments. Figure~\ref{fig:llama_logprobs} illustrates this challenge: querying the same prompt ("Hello my name is") 20 times with temperature 0 through OpenRouter reveals non-trivial variation in log probabilities across runs, even for identical inputs. Several factors may contribute to this non-determinism. First, hardware and software heterogeneity can introduce systematic variation through differences in how calculations are implemented~\citep{zhang2025hardware, schloegl2023}. Second, even when a user requests temperature zero, it is unclear whether the serving endpoint performs pure greedy decoding. While frameworks like vLLM explicitly disable dynamic sampling parameters (min-p, top-p, top-k) when temperature falls below a small epsilon~\citep{vllm-sampling-2024}, providers routing requests through platforms like OpenRouter may apply their own sampling pipelines, custom logits processors, or model-specific generation configurations that deviate from these defaults. More broadly, the growing ecosystem of dynamic, context-aware sampling methods, including min-p truncation~\citep{minh2025turning}, entropy-based adaptive sampling~\citep{entropix2024}, and adaptive attention temperature~\citep{velickovic2025softmax}, reflects a trend toward inference-time modifications that are increasingly difficult for external users to audit, making it uncertain whether any given API endpoint truly implements deterministic decoding even when explicitly requested.